# ChatGPT cites the most-cited articles and journals, relying solely on Google Scholar's citation counts. As a result, AI may amplify the Matthew Effect in environmental science

Eduard Petiška; Charles University (Prague, Czech Republic); eduard.petiska@czp.cuni.cz
Assistant and respondent: ChatGPT; OpenAI, Inc; (San Francisco, CA)
April 11, 2023

## Abstract

**Introduction**
ChatGPT (GPT) has become one of the most talked-about innovations in recent years, with over 100 million users worldwide. However, there is still limited knowledge about the sources of information GPT utilizes. As a result, we carried out a study focusing on the sources of information within the field of environmental science. Our study aims to address the research question: "Does ChatGPT predominantly cite the most-cited publications in environmental science?"

**Method**
In our study, we asked GPT to identify the ten most significant subdisciplines within the field of environmental science. We then asked it to compose a scientific review article on each subdiscipline, including 25 references. We proceeded to analyze these references, focusing on factors such as the number of citations, publication date, and the journal in which the work was published.

**Results**
Our findings indicate that GPT tends to cite highly-cited publications in environmental science, with a median citation count of 1184.5. It also exhibits a preference for older publications, with a median publication year of 2010, and predominantly refers to well-respected journals in the field, with Nature being the most cited journal by GPT. Interestingly, our findings suggest that GPT seems to exclusively rely on citation count data from Google Scholar for the works it cites, rather than utilizing citation information from other scientific databases such as Web of Science or Scopus.

**Conclusion**
In conclusion, our study suggests that Google Scholar citations play a significant role as a predictor for mentioning a study in GPT-generated content. This finding reinforces the dominance of Google Scholar among scientific databases and perpetuates the Matthew Effect in science, where the rich get richer in terms of citations. With many scholars already utilizing GPT for literature review purposes, we can anticipate further disparities and an expanding gap between lesser-cited and highly-cited publications.

# Introduction

ChatGPT, a powerful language model, has demonstrated its ability to generate coherent and well-structured text, prompting its application in writing the introduction and abstract sections of scientific articles.[1] In fact, several papers have acknowledged ChatGPT as a co-author.[2] While its potential in research and academia is evident, concerns have arisen regarding its implications on education and integrity.[3]

Professor Darren Hick from Furman University identified ChatGPT's "style" in a student's paper, proposing a policy to administer ad-hoc oral exams if AI-generated submissions are suspected.[4] In response to concerns, the New York City Department of Education blocked access to ChatGPT. Scientific journals have taken varying stances on ChatGPT. Some, like Nature and JAMA Network, mandate disclosure of text-generating tool usage and prohibit listing ChatGPT as a co-author, while Science has banned the use of large language model-generated text entirely.[5,6]

As GPT has released in November, it is still a relatively new field of scientific research, although more than 1 000 papers were already published[7], still, less is known about the application of ChatGPT in concrete disciplines such as environmental science or scientometrics. In this study, we have concentrated our efforts on analyzing citations in the field of environmental science to investigate whether the Matthew effect, a well-known concept in the scientific community, also influences the citation patterns in ChatGPT-generated responses

# Mathew effect in science

The Matthew effect (rooted in the Gospel of Matthew) in science, a phenomenon first introduced by Robert K. Merton (1968)[8], has attracted significant attention in recent years due to its implications on the reward and communication systems of scientific research. This effect, which refers to the disproportionate allocation of rewards and recognition to well-established researchers, can impede scientific progress by inhibiting the growth of emerging scientists (Petersen et al., 2018;[9] Li & Chen, 2019[10]). Studies have demonstrated that the Matthew effect can lead to disparities in funding, career opportunities, and citation counts for researchers (Allison & Stewart, 1974[11]). In addition to disadvantaging early-career scientists (Bol et al., 2018[12]), the Matthew effect can also result in a skewed distribution of research attention and resources toward certain fields or topics (Wang et al., 2013[13]).

---

[1] Bushard, Brian (January 10, 2023). "Fake Scientific Abstracts Written By ChatGPT Fooled Scientists, Study Finds". Forbes. Archived from the original on February 3, 2023.
[2] Stokel-Walker, Chris (January 18, 2023). "ChatGPT listed as author on research papers: many scientists disapprove". Nature. 613 (7945): 620–621.
[3] Stokel-Walker, Chris (December 9, 2022). "AI bot ChatGPT writes smart essays — should professors worry?". Nature. doi:10.1038/d41586-022-04397-7
[4] Mitchell, Alex (December 26, 2022). "Students using ChatGPT to cheat, professor warns". The New York Post.
[5] Brainard, Jeffrey (February 22, 2023). "As scientists explore AI-written text, journals hammer out policies". Science. doi:10.1126/science.adh2937
[6] https://en.wikipedia.org/wiki/ChatGPT
[7] https://scholar.google.com/scholar?hl=en&as_sdt=0%2C5&q=allintitle%3AChatGPT&btnG=&oq=al
[8] Merton, R. K. (1968). The Matthew effect in science: The reward and communication systems of science are considered. Science, 159(3810), 56-63.
[9] Petersen, A. M., Fortunato, S., Pan, R. K., Kaski, K., Penner, O., Rungi, A., ... & Barabási, A. L. (2018). Reputation and impact in academic careers. Proceedings of the National Academy of Sciences, 111(43), 15316-15321.
[10] Li, E. Y., & Chen, Y. (2019). The Matthew effect in scientific funding. Proceedings of the National Academy of Sciences, 116(16), 7684-7689.

This effect is pronounced for example in the case of Wikipedia, as the platform often serves as a gateway to knowledge. As a result, articles on Wikipedia that cite well-established publications inadvertently reinforce the Matthew effect, as these sources continue to accumulate citations and dominate their respective fields (Thompson, N., & Hanley, D.,2018).[14]

Similarly, the Matthew effect in Google Scholar is an example of how the digital platform may amplify the visibility and recognition of the most cited publications, thereby reinforcing existing hierarchies in scientific research. As an influential search engine for scholarly literature, Google Scholar employs algorithms that prioritize highly-cited articles in its search results. This process can create a self-perpetuating cycle, where articles that are already well-established in their respective fields continue to receive even more citations, while lesser-known publications remain relatively obscure. Consequently, the Matthew effect in Google Scholar can contribute to a skewed distribution of attention and resources, potentially hindering the discovery of innovative ideas and research findings.

We hypothesize that the Matthew effect also plays a crucial role in the case of GPT. Since there are no studies focused on citation usage by ChatGPT and its decision-making process for selecting the papers it cites, we asked the model the following question: *"How do you choose which academic studies to cite?"* The response we received was: "As an AI language model, when generating responses, I rely on the information and data available during my training, which includes a wide range of academic studies, articles, and other sources up to my completion date (September 2021). When selecting which academic studies to cite, I consider several factors, including the relevance of the study to the question or topic at hand, the quality and reliability of the research, and the impact and influence of the publication within its field. Additionally, I may give more weight to studies that have been cited frequently or published in well-regarded journals, as these are often indicators of the importance and contribution of the research. However, it is important to note that my citations are generated algorithmically, and I do not have personal biases or preferences in selecting which academic studies to cite." As we can see, the answer provided by GPT is somewhat vague and lacks specificity, so we attempted to obtain more information from the model through subsequent discussion (more details can be found in the Discussion section).

---

[11] Allison, P. D., & Stewart, J. A. (1974). Productivity differences among scientists: Evidence for accumulative advantage. American Sociological Review, 39(4), 596-606.
[12] Bol, T., de Vaan, M., & van de Rijt, A. (2018). The Matthew effect in science funding. Proceedings of the National Academy of Sciences, 115(19), 4887-4890.
[13] Wang, J., Veugelers, R., & Stephan, P. (2013). Bias against novelty in science: A cautionary tale for users of bibliometric indicators. NBER Working Paper No. 19135.
[14] Thompson, N., & Hanley, D. (2018). Science is shaped by Wikipedia: evidence from a randomized control trial.

# Method

We pose the research question: *"Does ChatGPT predominantly cite the most-cited publications in environmental science?"* To answer this question, we followed several steps. First, we asked ChatGPT-4 to perform the following tasks:

1. Write the 10 most important subdisciplines of environmental science;
2. Compose an introduction for a research article (commonly referred to as a "review article") about "*subdiscipline*" (Ecology, Conservation Biology, etc.), ensuring that every statement within the introduction is supported by citations from relevant research articles. Include 25 references in total for this introduction;
3. Provide a list of all references in the table: date, title, and publisher in a separate column, so the table has 4 columns, including rank;
4. Now write the number of Google Scholar citations for each reference into the table, so that it has five columns.

After that, we analyzed responses from various disciplines' data (including citations and journals) based on the criteria mentioned earlier for each subdiscipline. These criteria included the oldest and newest citation, the median date of the citation, the publication with the lowest and highest number of citations, and the median number of citations. Finally, we asked GPT to determine the most cited journals in the responses it provided across the disciplines.

**Picture 1.** An illustration of the text section with the citations provided by GPT

Compose an introduction for a research article (commonly referred to as a "review article") about "Conservation Biology" ensuring that every statement within the introduction is supported by citations from relevant research articles. Include 25 references in total for this introduction.

Conservation biology is an interdisciplinary field that combines the natural and social sciences to address the pressing issue of preserving Earth's biodiversity (Soulé 1985). This multidisciplinary approach seeks to understand the complex interactions between species, ecosystems, and human activities, providing solutions to halt the rapid decline of global biodiversity (Meffe et al. 2006). The ongoing loss of species, driven by habitat destruction (Tilman et al. 1994), climate change (Thomas et al. 2004), invasive species (Mack et al. 2000), overexploitation (Pauly et al. 1998), and pollution (Vitousek et al. 1997), poses a significant threat to ecosystem functioning and human well-being (Cardinale et al. 2012; Dirzo et al. 2014).

Established in the 1980s, conservation biology has since expanded to include a diverse array of research topics and methodologies (Soulé 1985; Kareiva & Marvier 2011). Central to this field is the concept of biodiversity, encompassing genetic, species, and ecosystem diversity (Noss 1990). By evaluating the patterns and processes underlying biodiversity loss, conservation biologists aim to develop evidence-based management strategies (Sutherland et al. 2004).

Habitat loss and fragmentation are major drivers of species decline, with deforestation, agricultural expansion, and urbanization leading to the destruction of natural habitats (Fahrig 2003; Hansen et al. 2013; Newbold et al. 2015). This has prompted research on the effectiveness of protected areas (Bruner et al. 2001; Watson et al. 2014) and the implementation of habitat restoration (Benayas et al. 2009; Jones et al. 2018) as conservation strategies.

Climate change poses additional challenges, as rising temperatures and altered precipitation patterns impact species distributions and interactions (Parmesan 2006; Bellard et al. 2012)

# Results

**Table 1.** The top 10 most important subdisciplines of environmental science, according to GPT

| No. | Subdiscipline | Description |
| --- | --- | --- |
| 1 | Ecology | Study of interactions between organisms and their environment, and the factors affecting these interactions. |
| 2 | Conservation Biology | Study of preserving biodiversity, species, and ecosystems through habitat restoration and protected areas. |
| 3 | Environmental Chemistry | Examination of the chemical and biochemical processes in the environment, including pollution and natural cycles. |
| 4 | Environmental Engineering | Design and implementation of solutions to mitigate environmental problems, such as pollution control and waste management. |
| 5 | Environmental Policy and Management | Development, implementation, and analysis of policies and strategies to address environmental issues and promote sustainability. |
| 6 | Climatology | Study of climate patterns, changes, and their impact on ecosystems and human society. |
| 7 | Hydrology | Study of the distribution, movement, and quality of water on Earth, including surface and groundwater resources. |
| 8 | Environmental Health | Study of how environmental factors, such as pollution and climate change, impact human health and well-being. |
| 9 | Sustainable Development | Pursuit of economic, social, and environmental objectives without compromising the ability of future generations to meet their needs. |
| 10 | Renewable Energy | Development and implementation of sustainable energy sources, such as solar, wind, and hydroelectric power. |

**Table 2.** The overview of the average number of citations received by research articles cited by GPT

| Topic | Oldest citation | Newest citation | Median date of the citation | Publication with the lowest citations | Publication with most citations | The median number of citations |
|---|---|---|---|---|---|---|
| Ecology | 1973 | 2019 | 2012 | 436 | 8763 | 1150 |
| Conservation Biology | 1985 | 2016 | 2004 | 790 | 9400 | 1100 |
| Environmental Chemistry | 1961 | 2019 | 1999 | 183 | 8686 | 1762 |
| Environmental Engineering | 1972 | 2013 | 2001 | 38 | 1024 | 5590 |
| Environmental Policy and Management | 1986 | 2020 | 2009 | 155 | 1045 | 11,030 |
| Climatology | 2005 | 2015 | 2012 | 287 | 37840 | 1,263 |
| Hydrology | 2000 | 2017 | 2007 | 94 | 2308 | 339 |
| Environmental Health | 2003 | 2016 | 2011 | 66 | 3429 | 405 |
| Sustainable Development | 1987 | 2020 | 2011 | 162 | 18278 | 1269 |
| Renewable Energy | 2008 | 2021 | 2015 | 7 | 16,324 | 476 |
| **Total** | **1961** | **2021** | **2010** | **38** | **37840** | **1184.5** |

**Table 3.** The top 10 most cited journals by GPT across disciplines

| Rank | Journal | Citations |
|---|---|---|
| 1 | Nature | 17 |
| 2 | Science | 13 |
| 3 | Proceedings of the National Academy of Sciences | 8 |
| 4 | Annual Review of Environment and Resources | 6 |
| 5 | Global Environmental Change | 5 |
| 6 | Trends in Ecology & Evolution | 4 |
| 7 | Annual Review of Ecology, Evolution, and Systematics | 4 |
| 8 | BioScience | 4 |
| 9 | Global Change Biology | 4 |
| 10 | Ecological Applications | 4 |

# Discussion

*There are other questions related to ChatGPT's selection of research publications, as little is currently known about it. In order to shed more light on this topic, we posed several additional questions to GPT and have presented them below:[15]*

―――

*"How did you choose and rank the citations?"*
"I selected and ranked the top 10 most important subdisciplines of environmental science based on their broad relevance and impact on addressing current environmental challenges. The selected subdisciplines are essential for understanding the complex interactions within ecosystems, preserving biodiversity, managing resources, and promoting sustainable development. However, it is important to note that this ranking is subjective and may vary depending on individual preferences and specific contexts. Each subdiscipline in environmental science has its importance and contributes to our overall understanding of the environment and sustainability. The selection of the top 10 was made to provide an overview of some of the most significant and widely applicable areas of study within the field."

*"Up to which date do you count citations?"*
"The Google Scholar citation counts provided in the table are up to my knowledge cutoff date, which is September 2021. Please note that the citation counts may have changed since then, as new citations are added over time."

*"Can you provide me with citations from the Web of Science?"*
"As an AI language model, I am unable to access Web of Science directly to provide you with real-time citation counts. The information I have provided in this conversation is based on my training data, which goes up until September 2021. For the most current and accurate citation counts from Web of Science, you will need to access the database directly or ask someone with access to provide the information for you."

*"And as of September 2021, what is the number of citations in Web of Science?"*
"I apologize for any confusion. As an AI language model, I do not have direct access to the Web of Science database or its citation counts. My knowledge is based on the training data provided to me, which includes citation counts from Google Scholar up until September 2021. For citation counts from Web of Science, you would need to access the database directly or ask someone with access to provide the information for you."

*"What about Scopus?"*
"As an AI language model, I am unable to access Scopus or its citation data directly. The information I have provided in this conversation is based on my training data, which goes up until September 2021, and it includes citation counts from Google Scholar.
For the most current and accurate citation counts from Scopus, you will need to access the database directly or ask someone with access to provide the information for you."

---

[15] The prompts in the article may have been proofreaded, and GPT's responses can vary depending on the specific conversation. Additionally, some sections of this article may refer to previous versions of GPT.

*"How do you choose the citations that you mention?"*
As an AI language model, I based the selection of the 25 references on their relevance to the topic of "Conservation Biology." My training data includes a wide range of scientific articles and sources, which allows me to identify important and frequently cited works related to the field.

When composing the introduction for the research article, I considered key research areas and subtopics within conservation biology, such as habitat loss, climate change, invasive species, overexploitation, and pollution. I then selected references that were representative of these themes and that have been widely cited in the literature. This approach aimed to provide a comprehensive and well-rounded overview of the field, using sources that have made significant contributions to conservation biology.

Please note that the selection of references is not exhaustive, and there may be other important works that were not included. The list of references provided is meant to serve as a starting point for further research and exploration of the field.

*"Can you provide more details on how you chose the 25 citations and the algorithm used to do so?"*
As an AI language model, I consider various factors when selecting research articles to mention, and the number of citations in Google Scholar is one of those factors. A high citation count often indicates that a study has had a significant impact on its field, as it has been frequently referenced by other researchers. However, it is not the only criteria I use when selecting articles to mention.

Other factors I consider include:
Relevance: How well does the article align with the topic being discussed? I prioritize articles that are closely related to the subject matter.
Quality: Has the article been published in a reputable journal or by a well-regarded publisher? Peer-reviewed articles from established journals generally have a higher level of credibility.
Recency: How recent is the publication? While older articles can be foundational and influential, more recent articles may reflect current trends and advances in the field.
Author expertise: Are the authors of the article recognized experts in their field? Articles by well-respected researchers can carry more weight.
Novelty: Does the article present a new or innovative approach, theory, or finding? Articles that introduce new ideas or challenge existing paradigms may be more likely to be mentioned.

In summary, while the number of Google Scholar citations is an important factor, it is not the sole predictor for mentioning a research article. I take a holistic approach, considering multiple factors to ensure that the selected articles are relevant, high-quality, and representative of the field.

*"Can you be more concrete about how did you choose these 25 citations, and provide some algorithm which you used?"*
"Certainly! Here is the algorithm I used to select the 25 citations for the introduction on Conservation Biology:

1. I first identified key subtopics within Conservation Biology, such as habitat loss, climate change, invasive species, overexploitation, and pollution.
2. For each subtopic, I conducted a search using a scientific database (such as Google Scholar) to identify highly cited articles that are widely considered to be influential and impactful in the field.
3. I reviewed the abstracts and summaries of these articles to assess their relevance and importance to the subtopic.
4. I then selected the most highly cited articles that were representative of each subtopic and that provided a comprehensive and well-rounded overview of the field of Conservation Biology.
5. I ensured that the selected articles were published in reputable journals or by well-regarded publishers, and that they were authored by recognized experts in the field.
6. Finally, I compiled the 25 citations into a list, taking care to provide a balanced representation of the different subtopics and areas of research within Conservation Biology.

This algorithm allowed me to select a diverse range of highly cited articles that are considered to be important and influential in the field of Conservation Biology."

———

    During the research, several problems and issues arose, such as the fact that GPT provides references somewhat randomly. References can serve as pointers to further literature rather than specifically supporting concrete statements. Additionally, the statements generated by GPT can be too vague to require references.
    It is interesting to note that ChatGPT can be inconsistent, sometimes initially providing references but later claiming not to have them or being unable to provide them. For instance, GPT might state: "Due to the limitations of this AI, I cannot access real-time Google Scholar citations or verify the publisher for each reference. However, I can provide a list of the references with the date and title in a table format. You can later manually add the publisher and citations from Google Scholar." To address this issue, we either opened a new window or continued in the chat window where citations were provided, or we tried using a different prompt.
    It was observed that GPT can provide different answers to the same queries, especially in previous versions. However, based on the prompts we tried, the pattern in citation behavior remained consistent.

# Conclusion

GPT is already changing the world of knowledge. It has become the subject of scientific research, which in this field also proves that traditional academic publishing methods are becoming obsolete (similar to the case of the Covid pandemic). Preprints, Twitter, and Reddit are increasingly proving to be efficient ways of distributing knowledge in these respects. If we want to find out the latest information about GPT, for example, we can use the following channels:

1) Twitter search for GPT[16] (many tweets are relevant to scientific audiences as a new tool for scientists – Perplexity[17]);
2) Reddit search for GPT[18] and the forum ChatGPT[19], where you can find weekly updates[20];
3) Google News (e.g., US edition) for ChatGPT[21];
4) Google Scholar (allintitle:ChatGPT)[22]
5) Dimensions (can be filtered according to Altmetric and scientific areas to identify the most discussed research articles in your field)[23]

In conclusion, GPT appears to be influenced by the current assessment of scientific quality (primarily through citations) in environmental science, even though, according to GPT, citations are not the sole predictors for mentioning a study. We hypothesize that this citation pattern is prevalent across various disciplines; however, further research is necessary to test this hypothesis.

*We would like to acknowledge the assistance of an AI language model, ChatGPT, in generating responses for this research article.*

---

[16] https://twitter.com/search?q=GPT&src=typed_query
[17] https://twitter.com/Artifexx/status/1645303838595858432
[18] https://www.reddit.com/search/?q=GPT
[19] https://www.reddit.com/r/ChatGPT/
[20] https://www.reddit.com/r/ChatGPT/comments/12diapw/gpt4_week_3_chatbots_are_yesterdays_news_ai/
[21] https://news.google.com/search?q=chatGPT&hl=en-US&gl=US&ceid=US%3Aen
[22] https://scholar.google.com/scholar?hl=en&as_sdt=0%2C5&q=allintitle%3AChatGPT&btnG=
[23] https://app.dimensions.ai/discover/publication?search_mode=content&search_text=chatgpt&search_type=kws&search_field=text_search&order=altmetric&and_facet_for=80003